\documentclass[oribibl]{llncs}

\usepackage{latexsym}
\usepackage{xspace,latexsym}
\usepackage{multicol,longtable}
\usepackage{pstricks}
\usepackage{pst-node}
\usepackage{amsmath}
\usepackage{amssymb}
\usepackage{color}
\usepackage{graphicx}
\usepackage{hhline}
\usepackage{epsfig}

\newcommand{\topsepspace}{\vspace{\topsep}}
\long\def\symbolfootnote[#1]#2{\begingroup%
\def\thefootnote{\fnsymbol{footnote}}\footnote[#1]{#2}\endgroup}

\newcommand{\ignore}[1]{}

\newcommand{\boxtheorem}{\hfill\ensuremath{\Box}}

\newcommand{\nit}[1]{{\it #1}}

\newcommand{\IC}{\nit{IC}}

\newcommand{\defproof}[2]{{\noindent\textbf{Proof of #1:\
}}#2 \boxtheorem \vspace{2mm}\\ \noindent
\ignore{\newline\topsepspace}}

\title{\vspace*{-4mm}
Complexity of Consistent Query Answering in Databases under
Cardinality-Based and Incremental Repair Semantics}
\author{{\bf Andrei Lopatenko}\thanks{Also: University
of Manchester, Department of Computer Science,
UK.}~~~~~~~~~~~~~~~~~~{\bf Leopoldo
Bertossi}\ignore{\thanks{Contact
author.}}\\
\hspace*{-9mm}Free University of Bozen-Bolzano~~~~~~~~~~Carleton University\\
Faculty of Computer Science ~~~~~~~~~~~School of Computer
Science\\
\hspace*{-3mm}Bozen-Bolzano, Italy. ~~~~~~~~~~~~~~~~~~~Ottawa, Canada.\\
\hspace*{5mm}\texttt{lopatenko@inf.unibz.it}
~~~~~~~~~~~\texttt{bertossi@scs.carleton.ca}}

\institute{}
\begin{document}
\pagestyle{plain}
\thispagestyle{empty}
\maketitle

%%%%%%%%%%%%%%%%%%%%%%%%%%%%%%%%%%%%%%%%%%%%%%%%%%%%%%%%%%%%%%%%%%%%%%%%

\vspace{-6mm}\begin{abstract}
  Consistent Query Answering (CQA) is the problem of computing
  from a database the answers to a query that are consistent with
  respect to certain integrity constraints that the database, as a
  whole, may fail to satisfy. Consistent answers have been
  characterized as those that are invariant under certain
  minimal forms of restoration of the database consistency.
  In this paper we investigate algorithmic and complexity theoretic
  issues of CQA under database repairs that minimally depart -wrt the cardinality
  of the symmetric difference- from the original database. Research on this kind of repairs has been
  suggested in the literature, but no systematic study had been done.
  Here we obtain first tight complexity bounds. We also address, considering for the first time a dynamic
  scenario for  CQA, the problem
  of incremental complexity of CQA, that naturally occurs when an originally
  consistent database becomes inconsistent after the execution of a sequence of
  update operations.  Tight bounds on incremental complexity are provided for various
  semantics under denial constraints, e.g.
  (a) minimum tuple-based repairs wrt
  cardinality, (b) minimal  tuple-based repairs wrt set inclusion, and (c) minimum numerical aggregation
  of attribute-based repairs. Fixed parameter tractability is also investigated in this dynamic
  context, where the size of the update sequence becomes the relevant parameter.
\end{abstract}

\section{Introduction}
Integrity constraints (ICs) capture the semantics of
data and are expected to be satisfied by a database in order to
keep its correspondence with the outside reality it is modelling.
However, it is often the case that IC satisfaction cannot be
guaranteed, and inconsistent database states are common, e.g. in
integrated databases, census databases, legacy data, etc.
\cite{bc-2003}.

Consistent Query Answering (CQA) is the problem of computing from
a database those answers to a query that are consistent with
respect to certain ICs, that the database as a whole may fail to
satisfy. Consistent answers have been characterized as those that
are invariant under minimal forms of restoration of the database
consistency \cite{abc-1999}. From this perspective, CQA is a form
of cautious reasoning from a database under integrity constraints.

The notion of minimal restoration of consistency was captured in
\cite{abc-1999} in terms of database {\em repairs}, i.e. new,
consistent  database instances that share the schema with the
original database, but differ from the latter by a {\em minimal
set of whole tuples under set inclusion}. In
\cite{bc-2003,fm-2005,abc-1999,clr-2003,abchrs-2003,cm-2005}
complexity bounds for CQA under this repair semantics have been
reported. However, less attention has received the semantics of
CQA based on ``cardinality-based  repairs" of the original
database that {\em minimize the number of whole database tuples
by} {\em which the instances differ}.

\begin{example}\label{ex:reps}
Consider a database schema $P(X,Y,Z)$ with the functional
dependency $X \rightarrow Y$. The inconsistent instance $D =
\{P(a,b,c), P(a,c,d), P(a,c,e)\}$, seen as a set of ground atoms,
has two repairs wrt set inclusion, namely $D_1=\{P(a,b,c)\}$ and
$D_2=\{P(a,c,d),$ $P(a,c,e)\}$, because the symmetric set
differences with the original instance, i.e. $\Delta(D,D_1),
\Delta(D,D_2)$, are minimal under set inclusion. However only
$D_2$ is a cardinality-based repair, because the cardinality
$|\Delta(D,D_2)|$ of the symmetric set difference becomes a
minimum. \boxtheorem
\end{example}
 In this paper we address the problem of obtaining
complexity bounds for CQA under the semantics given by
cardinality-based repairs, and we do this by introducing some
graph theoretic techniques and results that, apart from  being
interesting by themselves, have a wider applicability in the
context of CQA. Although research on cardinality- and tuple-based
repairs has been proposed and started before in the context of CQA
\cite{bc-2003}, no detailed analysis of their complexity theoretic
properties has been provided. In \cite{tplp} a brief illustration
was given of how to specify cardinality based repairs using logic
programs with weak cardinality constraints \cite{buccafurri2k} and
stable model semantics.

Our emphasis is on CQA, as opposed to  computing or checking
repairs. This is because we are usually not interested in
computing specific repairs (there are exceptions though, e.g. in
census-like data \cite{bbfl-2005}), but in characterizing and
computing consistent answers to queries. However, the repair
semantics we choose will have  an impact on CQA.

\begin{example}\label{ex:repsCQA} (example \ref{ex:reps} continued)
The query $P(x,y,x)?$ has $(a,c,d)$ and $(a,c,e)$ as consistent
answers under the cardinality semantics (the classic answers in
the only repair), but none under the set inclusion semantics
(there is no classic answer shared by the two repairs).
\boxtheorem
\end{example}
All the complexity bounds on CQA given  so far in the literature,
no matter what repair semantics is chosen, consider  {\em the
static case}: Given a snapshot of a database, a set of integrity
constraints, and a query, the problem is to find consistent query
answers. However, databases are essentially dynamic structures,
subject to update operations. In this paper we also  take into
account dynamic aspects of data, studying the complexity of CQA
when the consistency of a consistent database may be affected by
update actions.

\begin{example}\label{ex:inc} (examples \ref{ex:reps} and \ref{ex:repsCQA} continued)
The cardinality-based repair $D_2 = \{P(a,c,d), P(a,c,e)\}$ is
obviously consistent, however after the execution of the update
operation $\nit{insert}(P(a,f,d))$ it becomes inconsistent. In
this case, the only cardinality repair of $D_2 \cup \{P(a,f,d)\}$
is $D_2$ itself. So, CQA from $D_2 \cup \{P(a,f,d)\}$ amounts to
classic query answering from $D_2$. However, if we start from the
consistent instance $D' = \{P(a,c,d)\}$, executing the same update
operation leads to two cardinality repairs, namely $D'$, but also
$\{P(a,f,d)\}$, and now CQA from $D' \cup \{P(a,f,d)\}$ is
different from classic query answering from $D'$, because two
repairs have to be taken into account. \boxtheorem
\end{example}
In this case, it would be inefficient to compute a materialized
repair of the database or a consistent answer to the query from
scratch after every update. In this paper we investigate how a
pre-computed repair of the database at a previous step or the
original instance itself if it was already consistent can be used
to consistently answer queries after update operations. We provide
a unified approach to the study of the computational complexity of
incremental consistent query answering; and not only under
cardinality-based repairs, but also under other repair semantics,
like the classic {\em minimal set inclusion semantics} and the one
based on {\em minimization of changes of attribute values}
(attribute-based repairs) that have been already used in the
literature \cite{w-2003,franconi,bbfl-2005,flesca}.

Incremental algorithms have been developed to check integrity
constraint satisfaction \cite{Nicolas}. In a similar spirit we
find work on incremental database maintenance, i.e. integrity
constraint satisfaction, by means of compensating active rules
\cite{Widom}. However, to the best of our knowledge, incremental
CQA and incremental repair computation (under updates) have not
been treated before. We know, by data complexity theoretic
reasons,  that some first-order queries asking for consistent
answers cannot be expressed as first-order queries asking for
classic answers \cite{bc-2003,clr-2003,cm-2005,fm-2005}. However,
in the incremental, dynamic context consistent answers could be
expressed as classic answers to first-order queries. A similar
situation can be found in incremental evaluation of queries that,
statically, are not expressible in the query language at hand, but
incremental computation can be performed and expressed
\cite{lw-1999}.

Cardinality-based CQA as studied in this paper has interesting
properties that make it useful as a semantics for CQA. First of
all, as illustrated in Example \ref{ex:reps}, it is clearly the
case that every cardinality-based repair is also a set
inclusion-based repair, but not necessarily the other way around.
In consequence, the consistent query answers under cardinality
repairs form a superset of the consistent answers under the set
inclusion-based semantics. Actually, in situations where the
latter does not give any answers (c.f. Example \ref{ex:repsCQA}),
the former does return answers, which is good. They could be
further filtered out according to other criteria at a
post-processing phase. In extreme cases, when there is only one
database tuple in semantic conflict with the rest of a possible
large  set of other tuples, the existence of a set inclusion-based
repair containing the only conflicting tuple would easily lead to
an empty set of consistent answers. The cardinality-based
semantics would not allow such a repair.  (Example \ref{ex:comp}
below illustrates this situation.)

This feature of the cardinality-based repair semantics comes at a
 price. In Section \ref{sec:mcss} we prove that CQA has a higher
data complexity than the classic, set inclusion-based semantics,
actually $P^\nit{NP(log(n))}$-hard vs. $\nit{PTIME}$ for denial
constraints \cite{cm-2005}. On the other side, the
cardinality-based semantics has the interesting property that CQA,
a form of cautious (or \emph{certain}) reasoning (true in {\em
all} repairs) and its brave (or \emph{possible}) version, i.e.
true in {\em some} repair, are mutually poly-time reducible and
share the same complexity. This is established in Section
\ref{sec:mcss} by proving first some useful graph theoretic lemmas
about maximum independent sets. This result may not hold for
classic CQA.

Furthermore, we prove in Sections \ref{sec:incmcss} and
\ref{sec:other} that incremental CQA for conjunctive queries under
the cardinality-based semantics has a lower complexity than
incremental CQA for the classic semantics, actually  $\nit{PTIME}$
vs. $\nit{coNP}$-hard (in data complexity), which makes the
cardinality semantics more appealing in a dynamic setting.

As we just mentioned, incremental CQA under the cardinality semantics
is polynomial in data complexity, but naive algorithms are exponential
if the size of the update sequence is a part of the input and
the combined complexity is considered. In Section \ref{sec:incmcss} we study
 the  parameterized complexity \cite{downfel,grohe} of CQA,
actually for the incremental case and under cardinality-based semantics, where
the parameter is the size of the update sequence. We establish that the
problem is fixed parameter tractable by  providing a concrete parameterized
algorithm.

For comparison with the cardinality-based semantics, we obtain in
Section \ref{sec:other} new results on the static and incremental
complexity under the classic semantics (i.e. tuple- and set
inclusion-based distance) and the attribute-based semantics (i.e.
attribute value-based and minimization of attribute changes). We
prove that static CQA for the weighted version of the
attribute-based semantics and incremental CQA under the
attribute-based semantics become both $P^\nit{NP}$-hard in data.

We concentrate on relational databases and basically on the class
of denial integrity constraints, which includes most of the
constraints found in applications where inconsistencies naturally
arise, e.g. census-like databases \cite{bbfl-2005}, experimental
samples databases, biological databases, etc. Complexity results
refer to data complexity \cite{AHV95}. For complexity theoretic
definitions and classic results we refer to
\cite{papadimitriou94}, to \cite{AHV95} for foundations of
databases, and  to \cite{downfel} for parameterized complexity.

\section{Preliminaries}

A relational database $D$ can be identified with a finite set of
ground atoms of the form $R(\bar{t})$, where $R$ is a relation in
the database schema ${\cal D}$, and $\bar{t}$ is a finite sequence
of constants taken from the underlying database domain $\cal U$.
The ground atom $R(\bar{t})$ is also called a {\em database
tuple}.\footnote{We also use the term {\em tuple} to refer to a
finite sequence $\bar{t} = (c_1, \ldots, c_n)$ of constants of the
database domain $\cal U$, but a {\em database tuple} is a ground
atomic sentence with predicate in $\cal D$ (excluding built-ins
predicates, like comparisons).} The relational schema ${\cal D}$
determines a first-order language $L({\cal D})$ based on the
relation names, the elements of $\cal U$, and extra built-in
predicates.
 In the language $L({\cal D})$, integrity constraints are sentences, and queries are
 formulas, usually with free variables. We assume in this paper
 that sets $\IC$ of ICs are always consistent in the sense that
 they are simultaneously satisfiable as first-order
 sentences.
A database is {\em consistent} wrt to a given set of integrity
constraints $\IC$ if the sentences in $\IC$ are all true in $D$,
denoted $D \models \IC$. An answer to a query $Q(\bar{x})$, with
free variables $\bar{x}$, is a tuple  $\bar{t}$ that makes $Q$
true in $D$ when the variables in $\bar{x}$ are interpreted as the
corresponding values in $\bar{t}$, denoted $D \models Q[\bar{t}]$.

For a database $D$, possibly inconsistent with respect to  $\IC$,
the {\em consistent answers to a query $Q$ from $D$ wrt $\IC$} are
characterized as those answers that are invariant under all {\em
minimal} forms of restoration of consistency for $D$, where
minimality refers to some sort of  distance between the original
instance $D$ and alternative consistent instances.

\begin{definition}\label{def:rep}  \em For a database $D$,
  integrity constraints  $\IC$ and a partial order $\preceq_{D,{\cal S}}$ over
  databases depending on the original database $D$ and a repair
  semantics ${\cal S}$, a \emph{repair of $D$ wrt $\IC$ under ${\cal S}$} is an
  instance $D^\prime$ such that: (a) $D^\prime$ has the same schema
  and domain as $D$; (b) $D^\prime \models \IC$; and (c) there is no $D''$ satisfying (a) and (b), such
  that $D'' \prec_{D,{\cal S}} D^\prime$, i.e. $D'' \preceq_{D,{\cal S}} D^\prime$ and not
  $D^\prime \preceq_{D,{\cal S}} D''$.
  The set of all repairs is denoted with $\nit{Rep}(D, \IC,{\cal S})$. \boxtheorem
\end{definition}
The class $\nit{Rep}(D, \IC,{\cal S})$ depends upon  the semantics
$\cal S$, which determines the partial order  $\preceq$
 and the way repairs can be
obtained, e.g. by allowing both insertions and deletions of whole
database tuples \cite{abc-1999}, or deletions of them only
\cite{cm-2005}, or only changes of attribute values
\cite{w-2003,bbfl-2005,flesca}, etc. (c.f. Definition
\ref{def:dist}.)

\begin{definition}
\label{def:semantics} \em Let $D$ be a database, $\IC$
  a set of ICs, and $Q(\bar{x})$ a query.  (a) A ground tuple
  $\bar{t}$ is a \emph{consistent answer} to $Q$ wrt $\IC$ under semantics $\cal S$
  if for every $D' \in \nit{Rep}(D,\IC,\cal S)$,
  $D' \models Q[\bar{t}]$.~  (b)
  $\nit{Cqa}(Q,D,\IC,{\cal S})$ is the set of consistent answers to
  $Q$ in $D$ wrt $\IC$ under semantics ${\cal S}$. If $Q$ is a sentence (a boolean query),
  $\nit{Cqa}(Q,D,\IC,{\cal S}) := \{\nit{yes}\}$ when $D' \models Q$ for
  every $D' \in \nit{Rep}(D,\IC,\cal S)$, and
  $\nit{Cqa}(Q,D,\IC,{\cal S}) := \{\nit{no}\}$, otherwise. (c)
  $\nit{CQA}(Q,\IC,{\cal S}) := \{(D,\bar{t}) ~|~ \bar{t} \in
  \nit{Cqa}(Q,D,\IC,{\cal S})\}$, the \emph{decision problem of consistent
    query answering}.
\boxtheorem
\end{definition}
The decision problem of CQA just defined if for {\em the static
case}, in the sense that it only considers a snapshot of the
database. In the literature different notions of distance have
been considered, they give rise to different repair semantics. We
summarize  here the most common ones, those that will be
investigated in this work. In the following, $\Delta(D',D)$
denotes the symmetric difference $(D'\smallsetminus D) \cup
(D\smallsetminus D')$ of two database instances conceived both as
set of ground atoms.

\begin{definition} \label{def:dist} (minimality semantics)~ \em
(a) \emph{Minimal set inclusion semantics} (simply, S-repair
semantics) \cite{abc-1999}: $D' \preceq D'' :\Longleftrightarrow
\Delta(D',D) \subseteq \Delta(D'',D)$.~
    (b) \emph{Minimum cardinality set semantics} (C-repair semantics): $D' \preceq D'' :\Longleftrightarrow
    |\Delta(D',D)| \leq |\Delta(D'',D)|$. ~(c) \emph{Aggregate attribute
difference semantics} (A-repair semantics) minimizes a numerical
aggregation function over attribute
    changes throughout the database.
\boxtheorem
\end{definition}
Particular classes of  A-repairs can be found in
\cite{franconi,flesca}, where the aggregation function to be
minimized is the number of all attribute changes; and in
\cite{bbfl-2005}, where the function is the overall quadratic
difference obtained from the changes in numerical attributes
between the original database and the repair.

S-repairs and C-repairs are examples of {\em tuple-based repairs},
in the sense that consistency is restored by inserting and/or
deleting database tuples. A-repairs are {\em attribute-based
repairs}, under which database instances can be repaired by
changing attributes values in existing tuples only. Classes of
attribute-based repairs have been studied in
\cite{w-2003,franconi,bbfl-2005,flesca}. Another notion of
attribute-based repair, not explored so far, and not included in
Definition \ref{def:dist}, could minimize, set-theoretically, the
set of attribute changes, with priorities imposed on attributes.
We will consider other repair semantics later on (e.g. Definition
\ref{def:wc}) and particular cases of attribute-based repairs
(c.f. Section \ref{sec:attrib}).

It is easy to prove that every C-repair is an S-repair; and
consequently every consistent query answer under the S-semantics
is a
 consistent query answer under the C-semantics.
However, as Example \ref{ex:reps} shows, not every
 S-repair is an C-repair. In that  example,
 attribute-based repairs could be $\{P(a,c,c), P(a,c,d),$ $P(a,c,e)\}$,
 suggesting that we we made a mistake in the second argument of the first tuple, but
also $\{P(a,b,c), P(a,b,d),$ $ P(a,b,e)\}$. If the aggregate
function in Definition \ref{def:dist}(c) is the number of changes
in attribute values, the former would be a repair, but not the
latter. These  instances are neither S- nor C-repairs if the
changes of attribute values have to be simulated via deletions
followed by insertions.

Integrity constraints may be any first-order sentences written in
language $L({\cal D})$, but  most of our results refer to denial
constraints only.

\begin{definition}\em \label{def:denials}
 \emph{Denial constraints}  are integrity constraints of the form
    $\forall \bar{x} \neg(A_1\land\ldots\land A_m \wedge \gamma),$
where each $A_i$ is a database atom and $\gamma$ is a conjunction
of
     comparison atoms.
    \boxtheorem
\end{definition}
Notice that functional dependencies (FDs), e.g. $\forall x \forall
y \forall z \neg(R(x,y) \wedge R(x,z) \wedge y \neq z)$, are
binary denial constraints; and range constraints are one-database
atom denials. For denial ICs, tuple-based repairs are obtained by
tuple deletions only \cite{cm-2005}.

In this paper we concentrate on data complexity. We briefly recall
some of the complexity classes used in this paper. $\nit{FP}$ is a
class of functional problems associated with languages in the
class $P$ of decision problems that are solvable in polynomial
time. $P^\nit{NP}$  (or $\Delta_2^P$) is the class of decision
problems solvable in polynomial time by a machine that makes calls
to an $\nit{NP}$ oracle. $P^\nit{NP(log(n))}$ is similarly
defined, but the number of calls is logarithmic. It is not known
if $P^\nit{NP(log(n))}$ is strictly contained in $P^\nit{NP}$. The
functional class $\nit{FP}^\nit{NP(log(n))}$ is similarly defined.
The class $\Delta^P_3\!(\nit{log(n)})$ contains decision problems
that can be solved by a polynomial time machine that makes a
logarithmic number of calls to an oracle in $\Sigma^P_2$.

\section{Complexity of CQA under the C-Repair Semantics} \label{sec:mcss}

CQA under the minimal cardinality repair semantics (C-repair
semantics in Definition \ref{def:dist}(b)) has received less
attention in the literature than the same problem under the
S-repair semantics. An exception is \cite{tplp}, where C-repairs
were specified using logic programs with non-prioritized weak
constraints under the skeptical stable model semantics
\cite{buccafurri2k}. As a consequence, from results in
\cite{buccafurri2k} (c.f. also \cite{DLV03}), we obtain that an
upper bound on the data complexity of CQA under the C-repair
semantics is the class $\Delta^P_3\!(\nit{log(n)})$.
 In this section we investigate the static complexity
of tuple-based CQA under the C-repair semantics.

In \cite{abchrs-2003}, {\em conflict graphs} were first introduced
 to study the
  complexity of CQA for aggregate queries wrt FDs under the S-repair semantics. They
 have as vertices the database tuples and edges connect
 two tuples that simultaneously violate a FD. There is a one-to-one correspondence between
 S-repairs of the database and the set-theoretically
 maximal independent sets (simply called {\em maximal independent sets}) in the conflict graph.
 Similarly,  there is  a one-to-one correspondence
 between C-repairs and {\em maximum independent sets}  in the same graph (but now
 they are maximum in cardinality).

 Notice that, unless an IC forces a particular tuple
 not to belong to the database\footnote{We do not consider in this work such {\em non generic} ICs
 \cite{bc-2003}.}, every tuple in the original database belongs to
 some
 S-repair, but not necessarily to a C-repair (c.f. Example \ref{ex:reps}, where the
 tuple $P(a,b,c)$ does not belong to the only C-repair).
 In consequence, testing membership of vertices to some maximum
 independent set becomes a problem that is relevant to address.
 For  this purpose we will make good use of some graph
theoretic constructions and results about maximum independent sets
obtained from them, whose proofs  use a self-reducibility property
of independent sets that can be expressed as follows: For any
graph $G$ and vertex $v$, every maximum independent set that
contains $v$ (meaning maximum among the independent sets that
contain $v$) consists of vertex $v$ together with a maximum
independent set of the graph $G'$ that is obtained from $G$ by
deleting all vertices  adjacent to $v$.

\begin{lemma}\label{lem:graphExt}\em
Given a graph $G$ and a vertex $v$ in it, a graph $G'$ that
extends $G$ can be constructed in polynomial time in the size of
$G$, such that there is a maximum independent set $I$ of $G$
containing $v$ iff $v$ belongs to every maximum independent set of
$G'$ iff the sizes of maximum independent sets in $G$ and $G'$
differ by one. \boxtheorem
\end{lemma}
Actually the graph $G'$ in this lemma can be obtained by adding a
new vertex $v'$ that is connected only to the neighbors of $v$.
Conversely, the following holds

\begin{lemma}\label{lem:all2some} \em
 For every graph $G$ and vertex $v$ there is a graph $G'$ that can
 be  constructed in polynomial time in the size of $G$, such that
 $v$ belongs to all maximum independent sets of $G$ iff $v$
 belongs to some maximum independent set of $G'$. \boxtheorem
 \end{lemma}
From the lemmas and the membership to $\nit{FP}^\nit{NP(log(n))}$
of computing the size of a maximum clique in a graph
\cite{krentel}, we obtain

 \begin{proposition}\label{lem:dec-graphNPlogn} \em The problems of deciding
  for a vertex in a graph if it belongs to some maximum
  independent set and
  if it belongs to all maximum
  independent set are both in $P^\nit{NP(log(n))}$. \boxtheorem
\end{proposition}
Since a ground atomic query is consistently true when it belongs,
as a database tuple, i.e. as a vertex in the conflict graph, to
all the maximum independent sets of the conflict graph, we obtain

\begin{corollary}\label{prop:memb}
\em For functional dependencies and ground atomic queries,  CQA
under the C-repair semantics belongs to $P^\nit{NP(log(n))}$.
\boxtheorem
\end{corollary}
Considering the maximum independent sets (or the C-repairs) as a
collection of possible worlds,
 the previous lemmas show a close connection between the {\em
 certain}
 and {\em possible} C-semantics, that sanctions
 something as true if it is true in  {\em every} (the default for CQA), resp. {\em some}
 possible world. CQA under these semantics and functional dependencies are polynomially reducible
 to each other, and also share the same complexity.

Using this result, it is possible to extend
 Corollary \ref{prop:memb} to negative atomic queries because $P^\nit{NP(log(n))}$ is closed
 under complement: Notice that a
vertex does not belongs to any maximum independent sets means that
the {\em certain answer} to the corresponding negated query is
{\em yes} and the {\em possible answer} to the corresponding
atomic query is {\em no} (or better, {\em false} in the sense that
it is false in every C-repair). On the other side, that there is a
maximum independent set to which a vertex does not belong means
that the {\em possible answer} to the corresponding negated query
is {\em yes} and the {\em certain answer} to the positive query is
{\em no}. Corollary \ref{prop:memb} also holds for queries that
are conjunctions of atoms.

 The next
result shows that graphs with their maximum independent sets can
be uniformly encoded as database repair problems under the
C-semantics.

 \begin{proposition}\label{lem:everyGraph}\em
 There is a fixed database schema ${\cal D}$ and a denial constraint $\varphi$ in
 $L({\cal D})$, such that for every graph $G$, there is an instance
 $D$ over ${\cal D}$, whose C-repairs wrt $\varphi$ are in
 one-to-one correspondence with the maximum independent sets of
 $G$. Furthermore, $D$ can be built in polynomial time in the size
 of $G$. \boxtheorem
 \end{proposition}
 This proposition  is a representation result, of
the maximum independent sets of a graph as the C-repairs of an
inconsistent database wrt a denial constraint. This is
interesting, because conflict graphs for databases wrt denial
constraints are actually {\em conflict hypergraphs} \cite{cm-2005}
that have as vertices the database tuples, and as hyperedges the
(set theoretically minimal) collections of tuples that
simultaneously violate one of the denial constraints.

The correspondence for conflict graphs between repairs and
independent sets -maximum or maximal  depending on the semantics-
still holds for hypergraphs, where an independent set in an
hypergraph is a set of vertices that does not contain any
hyperedges \cite{cm-2005}.  Lemmas \ref{lem:graphExt} and
\ref{lem:all2some} and Proposition \ref{lem:dec-graphNPlogn} still
hold for hypergraphs, and in consequence the polynomial time
mutual reducibility between the {\em certain} and {\em possible}
semantics for CQA still holds for denial constraints and ground
atomic queries.

From Proposition \ref{lem:everyGraph} and the
$P^\nit{NP(log(n))}$-completeness of determining the size of a
maximum clique \cite{krentel}, we obtain
\begin{corollary} \label{cor:size} \em
Determining the size of a C-repair for denial constraints is
complete for $P^\nit{NP(log(n))}$. \boxtheorem
\end{corollary}
Using the hypergraph representation of C-repairs, it is possible
to generalize Corollary \ref{prop:memb} to  the case of denial
constraints and queries that are conjunctions of atoms.

\begin{proposition}\label{prop:membDCs} \em
For denial constraints and non-existentially quantified
conjunctive queries, CQA under the C-repair semantics belongs to
$P^\nit{NP(log(n))}$. \boxtheorem
\end{proposition}
This result can be generalized to conjunctive queries containing
negation, but no quantifiers. For example, the query $Q = A_1
\land \cdots \wedge A_i \land \neg A_{i+1} \cdots \land \neg A_k$,
is true iff each positive conjunct  is contained in every maximum
independent set, and each atom preceded by a negation is not
contained in any maximum independent sets. Deciding if a database
tuple $A$ is not contained in any maximum independent sets of a
graph $G$ is in $\nit{FP}^{\nit{NP(log}(n))}$, because it is the
complement to the problem of deciding if a database tuple $A$ is
contained in some maximum independent sets of a graph, and
$\nit{FP}^{\nit{NP(log}(n))}$ is closed under complement.

\begin{multicols}{2}
In order to obtain hardness for CQA under the C-repair semantics,
we need a useful graph theoretic construction, the block
$B_k(G,{\bf t})$ (c.f. Figure 1), consisting of two copies
$G_1,G_2$ of $G$, and two internally disconnected subgraphs $I_k,
I_{k+1}$, with $k$ and $k+1$ vertices, resp. Every vertex in $G$
($G'$) is connected to every vertex in $I_k$ (resp. $I_{k+1}$).

\phantom{MMMMMMM}

\vspace*{3.5cm} \psset{xunit=0.5cm,yunit=0.5cm} \hspace*{10mm}
\begin{pspicture}{0,0}{10,8}[h]
\psellipse[linestyle=dashed, fillstyle=none](1,2)(1,2)
\psellipse[linestyle=dashed, fillstyle=none](1,8)(1,2)
\psellipse[linestyle=dashed, fillstyle=none](5,8)(1,2)
\psellipse[linestyle=dashed, fillstyle=none](5,2)(1,2)
\qdisk(8.5,8){2pt} \qdisk(8.5,2){2pt}
\qdisk(5,8){2pt} \qdisk(5.2,8.4){2pt} \qdisk(4.8,7.5){2pt}
\qdisk(4.8,9.6){2pt} \qdisk(4.7,6.3){2pt} \qdisk(5,7){2pt}
\qdisk(5.5,7,4){2pt} \qdisk(4.5,8.3){2pt}
\qdisk(5,2){2pt} \qdisk(5.2,2.4){2pt} \qdisk(4.8,1.5){2pt}
\qdisk(4.8,3.6){2pt} \qdisk(4.7,0.3){2pt} \qdisk(5,1){2pt}
\qdisk(5.5,1,4){2pt} \qdisk(4.5,2.3){2pt}
\psline{-}(8.5,8)(8.5,2)
\psline(8.5,8)(5,8) \psline(8.5,8)(5.2,8.4)
\psline(8.5,8)(4.8,7.5) \psline(8.5,8)(4.8,9.6)
\psline(8.5,8)(4.7,6.3) \psline(8.5,8)(5,7)
\psline(8.5,8)(5.5,7,4) \psline(8.5,8)(4.5,8.3)

\psline(8.5,2)(5,2) \psline(8.5,2)(5.2,2.4)
\psline(8.5,2)(4.8,1.5) \psline(8.5,2)(4.8,3.6)
\psline(8.5,2)(4.7,0.3) \psline(8.5,2)(5,1)
\psline(8.5,2)(5.5,1,4) \psline(8.5,2)(4.5,2.3)
\psline{-}(5.2,8.4)(1.5,7.6) \psline{-}(5.2,8.4)(0.7,8.4)
\psline{-}(5.2,8.4)(1.1,9.3) \psline{-}(5.2,8.4)(0.8,7.2)
\psline{-}(4.7,6.3)(1.5,7.6) \psline{-}(4.7,6.3)(0.7,8.4)
\psline{-}(4.7,6.3)(1.1,9.3) \psline{-}(4.7,6.3)(0.8,7.2)
\psline{-}(4.8,9.6)(1.5,7.6) \psline{-}(4.8,9.6)(0.7,8.4)
\psline{-}(4.8,9.6)(1.1,9.3) \psline{-}(4.8,9.6)(0.8,7.2)
\psline{-}(5,7.5)(1.5,7.6) \psline{-}(5,7.5)(0.7,8.4)
\psline{-}(5,7.5)(1.1,9.3) \psline{-}(5,7.5)(0.8,7.2)
\psline{-}(5.2,2.4)(1.5,1.6) \psline{-}(5.2,2.4)(0.7,2.4)
\psline{-}(5.2,2.4)(1.1,3.3) \psline{-}(5.2,2.4)(0.8,1.2)
\psline{-}(4.7,0.3)(1.5,1.6) \psline{-}(4.7,0.3)(0.7,2.4)
\psline{-}(4.7,0.3)(1.1,3.3) \psline{-}(4.7,0.3)(0.8,1.2)
\psline{-}(4.8,3.6)(1.5,1.6) \psline{-}(4.8,3.6)(0.7,2.4)
\psline{-}(4.8,3.6)(1.1,3.3) \psline{-}(4.8,3.6)(0.8,1.2)
\psline{-}(5,1.5)(1.5,1.6) \psline{-}(5,1.5)(0.7,2.4)
\psline{-}(5,1.5)(1.1,3.3) \psline{-}(5,1.5)(0.8,1.2)
\psellipse[linestyle=dashed, fillstyle=none](1,2)(1,2)
\uput[u](8.8,8){\large{\textbf{t}}}
\uput[u](8.8,2){\large{\textbf{b}}}
\uput[u](5.5,9.7){\large{\textbf{$I_k$}}}
\uput[u](5.5,3.7){\large{\textbf{$I_{k+1}$}}}
\uput[u](1.5,9.7){\large{\textbf{$G_1$}}}
\uput[u](1.5,3.7){\large{\textbf{$G_2$}}}
\end{pspicture}

\vspace{12mm} {\small \bf \hspace*{5mm}Figure 1. ~~The block
$B_k(G,t)$}
\end{multicols}

\begin{lemma} \label{lem:block} \em ({\em the block
construction})~ Given a  graph $G$ and a number $k$, there exists
a graph $B_k(G,{\bf t})$, where ${\bf t}$ is a distinguished
vertex in it, such that ${\bf t}$ belongs to all maximum
independent sets of $B_k(G,{\bf t})$ iff the cardinality of a
maximum independent set of $G$ is equal to $k$. ~$B_k(G,{\bf t})$
can be computed in polynomial time in the size of $G$. \boxtheorem
\end{lemma}

\begin{proposition} \label{prop:allsets} \em
Deciding if a vertex belongs to all maximum independent sets of a
graph is $P^{\nit{NP(log}( n))}$-hard. \boxtheorem
\end{proposition}
This result can be proved by reduction from the following
$P^{\nit{NP}(\nit{log}(n))}$-complete decision problem
\cite{krentel}: Given a graph $G$ and an integer $k$, is the size
of a maximum clique in $G$ equivalent to $0~ \nit{mod}~ k$? $G$ is
reduced to a graph that is built by combining a number of versions
of the block construction in Figure 1. Now, using Proposition
\ref{lem:everyGraph}, the graph constructed for the reduction in
Proposition \ref{prop:allsets} can be represented as a database
consistency problem, and in this way we~obtain

\begin{theorem} \label{teo:hardness} \em
For denial constraints, CQA for ground atomic  queries under the
C-repair semantics is $P^{\nit{NP}(\nit{log}( n))}$-complete.
\boxtheorem
\end{theorem}
This theorem is interesting, because CQA for denial constraints,
but S-semantics is in \textit{PTIME} for arbitrary ground atomic
queries \cite{cm-2005}; and also because query answering under the
S-semantics in the context of belief revision/update is more
complex than the same problem for C-semantics (assuming the
polynomial hierarchy does not collapse); more precisely Winslett's
framework \cite{winslett} (based on set inclusion) is
$\Pi^P_2$-complete, while Dalal's \cite{dalal} (based on set
cardinality) is $P^\nit{NP(log(n))}$-complete \cite{eg-92}.
Connections between CQA and belief revision were already
established in \cite{abc-1999}. Notice that our complexity results
do not follow, at least not straightforwardly from the results for
belief revision presented in \cite{eg-92}. They apply in the
propositional setting, in combined complexity (as opposed to data
complexity), and the revision formulas (in our case the
constraints) and the query do not necessarily satisfy our
conditions.

 Now we consider a weighted version of
the C-semantics. Under denial constraints, this means that it may
be more costly to remove certain tuples than others to restore
consistency.

\begin{definition}(we\!C-semantics)~ \label{def:wc}\em
Assume that every database tuple $R(\bar{t})$ in $D$ has an
associated numerical cost $w(R(\bar{t}))$.  $D'$ is a repair of
$D$ under the {\em weighted minimum cardinality set semantics} if
the order relation used in Definition \ref{def:rep} is given by
$D_1 \preceq_{D,w} D_2 :\Longleftrightarrow |D \triangle D_1|_w
\leq |D \triangle D_2|_w$, where $|S|_w$ for a set of database
tuples $S$ is the sum of the weights of the elements of $S$.
\boxtheorem
\end{definition}
This semantics is a generalization of the C-semantics, that can be
obtained by defining all the weights as $1$; and as such, it is
still tuple-based, actually tuple deletion-based for denial
constraints.

\begin{proposition}\label{theo:weightedpro} \em CQA for ground atomic
queries wrt denial constraints under the weC-repair semantics
belongs to $P^\nit{N\!P}$. \boxtheorem
\end{proposition}

\section{Incremental Complexity of CQA}\label{sec:incr}

We will consider the following problem of  \emph{incremental CQA}:
Assume that we have a consistent database $D$ wrt to certain
integrity constraints. After an update sequence $U$ on $D$
composed of update operations of any of the forms
$\nit{insert}(R(\bar{t})), ~\nit{delete}(R(\bar{t}))$, ~meaning
insert/delete tuple $R(\bar{t})$ into/from $D$, ~or \newline
$\nit{change}(R(\bar{t}),A,a)$,  for changing value of attribute
$A$ in $R(\bar{t})$ to $a$, with $a \in {\cal U}$, we may obtain
an inconsistent database. We are interested in  whether we can
find consistent query answers from the updated inconsistent
database more efficiently by taking into account the previous
consistent database state. We will see that for a few particular
cases of the consistent query answering problem, the knowledge of
the previous state may significantly simplify consistent query
answering, while in other cases the worst-case computational
complexity of query answering is the same as in  the classic,
static scenario, where no updates are considered.

\begin{definition} \em For a set of integrity constraints $\IC$, a
  database $D$ that is consistent wrt $\IC$,
  and a sequence $U$ of one-tuple database update operations $U_1,\ldots,U_m$,
\emph{incremental consistent query answering} for query $Q$ is CQA
for $Q$ wrt $\IC$  from  the instance
  $U(D)$ that results
from applying $U$ to $D$. The complexity of incremental CQA is
measured wrt the size of $D$. \boxtheorem
\end{definition}
\vspace{-2mm} In order not to add complexity on the sole basis of
the length of the update sequence, we will usually assume that $m$
is small in comparison to the size of the underlying database $D$,
say $m < c \cdot |D|$. We are in general interested in data
complexity, i.e. wrt $|D|$. However, in Section \ref{sec:incmcss}
we consider parameterized complexity, where the role of parameter
$m$ is considered. Furthermore, we consider an update sequence $U$
as atomic in the sense that it is completely executed or not. In
particular, this allows us to concentrate on ``minimized" versions
of update sequences, e.g. containing only insertions and/or
attribute changes when dealing with denial constraints, because
deletions do not have any effects on them.

A notion of incremental complexity has been introduced in
\cite{msvt-1994}, and also in \cite{imm-99} under the name of {\em
dynamic complexity}. However, our notion is different. In those
papers, the instance that is updated can be arbitrary, and the
question is about the complexity for the updated version when
information about the previous instance can be used. In our case,
we are assuming that the initial database is consistent, and then
the problem of finding its repairs is trivial (the only repair is
the database itself) and CQA is easy (just query the database as
usual). However, the same problems for the updated database are
not necessarily trivial or easy. Furthermore, as opposed to
\cite{msvt-1994,imm-99}, where new incremental or dynamic
complexity classes are introduced, we appeal to those classic
complexity classes found at a low level in the polynomial
hierarchy, which are applied to data complexity, relative to the
size of the initial database.

\subsection{Incremental complexity: C-repair semantics}
\label{sec:incmcss}

In this section we study the computational complexity of CQA under
the C-semantics over an inconsistent database that is obtained
trough a short sequence of update operations on a consistent
database.

\begin{proposition}\label{prop:incr-dens}
\em Under the C-semantics, incremental
  CQA  for first-order boolean queries, denial constraints, and a sequence of atomic updates
  $U\!: U_1,$ $\ldots, U_m$ applied to a database $D$
   is in \textit{PTIME} in the size of $D$. \boxtheorem
  \end{proposition}
As we saw in Theorem \ref{teo:hardness}, static CQA for denial
constraints under the C-semantics is a hard problem, in contrast
to incremental CQA for the same class of queries and constraints.
For the latter problem, an upper bound of of $O(m \cdot n^m)$ can
be obtained (c.f. proof of Proposition \ref{prop:incr-dens}), that
is polynomial in the size $n$ of the initial database, but
exponential in $m$. So, the problem is tractable in data
complexity, but the size of the update sequence is in the exponent
of $n$. We are interested in determining if a query can be
consistently answered in $O(f(m) \times n^c)$, where $c$ is a
constant and $f(m)$ is a function which depends only on $m$, and
by doing so, to isolate the complexity introduced through the
update.

The area of parameterized complexity (or fixed parameter
tractability) \cite{downfel,grohe,papyan-1999} provides the right
tools to attack this problem. A decision problem with inputs of
the form $(I,p)$, where $p$ ~is a distinguished parameter of the
input, is {\em fixed parameter tractable}, and by definition
belongs to the class $\nit{FPT}$ \cite{downfel}, if it can be
solved in time $O(f(|p|) \cdot |I|^c)$, where $c$ and the hidden
constant do not depend on $|p|$ or $|I|$ and $f$ does not depend
on $|I|$.

\begin{definition} (parameterized CQA) ~\em Given a query $Q$,
a set of ICs $\IC$, and a ground tuple $\bar{t}$, \emph{the
parameterized complexity of CQA} is the complexity of the decision
problem $\nit{CQA}^p( Q,\IC, \bar{t}) := \{(D,U) ~|~ D \mbox{ is
an}$ $\mbox{instance, } U \mbox{ an update se-}$ $\mbox{quence },$
$\bar{t}~ \mbox{ is consistent answer to } Q \mbox{ in } U(D)\}$,
whose parameter is $U$, and the consistency of an answer refers to
the C-repairs of $U(D)$. \boxtheorem
\end{definition}
We fixed $Q, \IC$ and $\bar{t}$ in the problem definition because,
except for the parameter $U$, we are interested in data
complexity. We emphasize that we are considering the parameterized
version of incremental CQA, and not of CQA in general.

\begin{proposition}\label{prop:fpt} \em
Incremental CQA for atomic ground queries and functional
dependencies under the C-repair semantics is in $\nit{FPT}$, being
the parameter involved the size $m$ of the update sequence.
\boxtheorem
\end{proposition}
The \emph{vertex cover problem} belongs to the class $\nit{FPT}$,
i.e. there is a polynomial time parameterized algorithm that
solves it, say $\nit{VC(G,k)}$, that determines if graph $G$ has a
vertex cover of size no bigger than $k$ \cite{downfel}, e.g. there
is one that runs in time $O(1.2852^k + k \cdot n)$, being $n$ the
size of $G$ \cite{chen}.

The algorithm whose existence is claimed in Proposition
\ref{prop:fpt} is essentially as follows (c.f. its proof for
details): Let $G$ be the conflict graph associated to the database
obtained after the insertion of $m$ tuples. By binary search,
calling each time $\nit{VC}(G,\_)$, it is possible to determine
the size of a minimum vertex cover for $G$. This gives us the
minimum number of tuples that have to be removed in order to
restore consistency; and can be done in time $O(\nit{log}(m) \cdot
(1.2852^m + m \cdot n))$, where $n$ is the size of the original
database. In order to determine if a tuple $R(\bar{t})$ belongs to
every maximum independent set, i.e. if it is consistently true,
compute the size of a minimum vertex cover for $G \smallsetminus
\{R(\bar{t})\}$. The two numbers are the same iff the answer is
$\nit{yes}$. The total time is still $O(\nit{log}(m) \cdot
(1.2852^m + m \cdot n)))$, which is linear in the size of the
original database.

The same algorithm applies if, in addition to tuple insertions, we
also have changes of attribute values in the update part; of
course, still under the C-repair semantics.

Having established the fixed parameter tractability of CQA, it
becomes relevant to find better parameterized algorithms to solve
this problem. The proof of Proposition \ref{prop:fpt} uses the
membership to $\nit{FPT}$ of the {\em vertex cover problem} for
graphs \cite{downfel}. That is why we restricted ourselves to
functional dependencies, that are associated to conflict graphs.
However, the result can be extended to denials constraints. In
fact, in this case we have conflict hypergraphs, but the maximum
size of an hyperedge is the maximum number of database atoms in a
denial constraint, which is determined by the fixed database
schema. If this number is $d$, then we are in the presence of the
so-called \emph{d-hitting set problem}, consisting in finding the
size of a minimum hitting set for an hypergraph with hyperedges
bounded in size by $d$. This problem is in $\nit{FPT}$
\cite{niedermeier}.

\begin{theorem} \label{teo:fpt} \em
Incremental CQA for atomic ground queries and denial constraints
under the C-repair semantics is in the class $\nit{FPT}$, being
the parameter involved the size of the update sequence.
\boxtheorem
\end{theorem}
The membership to $\nit{FPT}$ can be extended to the incremental
CQA under the {\em possible semantics} that sanctions as true what
is true of some C-repair. This is due to the reduction of this
semantics to the {\em certain semantics} exhibited in Section
\ref{sec:mcss}, which requires the introduction of only a few
extra vertices (this also holds for denial constraints and their
hypergraphs).

In a different direction, incremental CQA, considered as a
parameterized problem in the size of the update, becomes
$\nit{MONOTONE}~W[1]$-hard \cite{downfel}, where the class
$\nit{MONOTONE}~W[1]$ is defined as $W[1]$, but in terms of
monotone circuits \cite{df-1995}. This result uses a uniform and
parameterized reduction from the $\nit{MONOTONE~W[1]}$-hard
problem \cite{df-1995} \emph{WEIGHTED MONOTONE 3CNF SAT}:

\begin{proposition}\label{lem:param-compl-incr} \em The
parameterized complexity of CQA wrt denial constraints under
C-repair semantics is
 $\nit{MONOTONE}~W[1]$-hard. \boxtheorem
\end{proposition}
Since $\nit{MONOTONE}~W[1]$ coincides with the class $\nit{FPT}$
\cite{downfel}, we conclude that incremental CQA under this
setting is $\nit{FPT}$-hard.

\section{Incremental Complexity of CQA: Other Semantics}
\label{sec:other}

In this section we consider the S-repair semantics  based on set
difference of tuples, and the A-repair semantics  based on changes
of attribute values  (c.f. Definition \ref{def:dist}).

\subsection{S-repair semantics}
Incremental CQA for non-quantified conjunctive queries under
denial constraints belongs to $\nit{PTIME}$, which can be
established by applying the algorithm in \cite{cm-2005} for the
static case. However,  for quantified conjunctive queries the
incremental CQA for the S-repair semantics is not in $\nit{PTIME}$
anymore, which contrast to incremental CQA for the C-repair
semantics (c.f. Proposition \ref{prop:incr-dens}). In fact, by
reduction from static
  CQA for conjunctive queries and denial ICs under the S-repair semantics,
  which is \textit{coNP}-hard
  \cite{cm-2005}, we obtain

\begin{proposition}\label{lem:incr-icd-tds-msd} \em
Under the S-repair semantics,
  incremental CQA for conjunctive queries and denial constraints is
  \textit{coNP}-hard. \boxtheorem
\end{proposition}
Despite the fact that static CQA is harder for the C-repair
semantics than for the S-repair semantics for denial constraints
($P^{\nit{NP(log}(n))}$ vs. $\nit{coNP}$-hard), incremental CQA
under the S-repair semantics is harder than the same problem under
the C-repair semantics. The reason is that for the C-repair
semantics the cost of a repair cannot exceed the size of an
update, whereas for the S-repair semantics the cost of a repair
may be unbounded wrt the size of an update.

\begin{example}\label{ex:comp}
Consider a schema $R(\cdot), S(\cdot)$ with the denial constraint
$\forall x \forall y\neg (R(x) \wedge S(y))$; and the consistent
database $D = \{R(1), \ldots, R(n)\}$, with an empty table for
$S$. After the update $U= insert(S(0))$, the database becomes
inconsistent, and the S-repairs are $\{R(1), \ldots, R(n)\}$ and
$\{S(0)\}$. However, only the former is a C-repair, and is at a
distance $1$ from the original instance, i.e. as the size of the
update. However, the second S-repair is at a distance $n$.
\boxtheorem
\end{example}

\subsection{A-repair semantics}
\label{sec:attrib}

Before addressing the problem of incremental complexity, we give a
complexity lower bound for the weighted version of static CQA for
A-repairs. For this case, we need a weight function $w$ that sends
triples of the form $(R(\bar{t}),A,\nit{newValue})$, where
$R(\bar{t})$ is a database tuple stored in the database, $A$ is an
attribute of $R$, and $\nit{newValue}$ is a new value for $A$ in
$R(\bar{t})$, to numerical values. The \emph{weighted A-repair
semantics} (wA-repair semantics) is just a particular case of
Definition \ref{def:dist}(c), where the distance is given by an
aggregation function $g$ applied to the set of numbers
$\{w(R(\bar{t}),A,\nit{newValue})~|~ R(\bar{t}) \in D\}$.

Typically, $g$~ is the sum, and the weights are
$w(R(\bar{t}),A,\nit{newValue}) = 1$ if $R(\bar{t})[A]$ is
different from $\nit{newValue}$, and $0$ otherwise, where
$R(\bar{t})[A]$ is the projection of database tuple $R(\bar{t})$
on attribute $A$. That is, just the number of changes is counted.
However, in \cite{bbfl-2005}, $g$ is still the sum, but the weight
function is given by $w(R(\bar{t}),A,\nit{newValue}) =
\alpha_{\!A} \!\cdot \!(R(\bar{t})[A] - \nit{newValue})^2$, where
$\alpha_A$ is a coefficient introduced to capture the  relative
importance of attribute $A$ or scale factors.

\begin{theorem}\label{theo:wmccs} \em Static CQA for ground
atomic queries and denial constraints under the wA-repair
semantics is $P^\nit{NP}$-hard. \boxtheorem
\end{theorem}
To obtain complexity lower bounds for incremental CQA under this
repair semantics, we need first a technical result

\begin{lemma}\label{lem:3col4reg} \em For any planar graph $G$
with vertices of degree at most 4, there exists a regular graph
$G'$ of degree 4 that is 4-colorable, such that $G'$ is
3-colorable iff $G$ is 3-colorable, and $G'$ can be built in
polynomial time in the size of $G$. \boxtheorem
\end{lemma}
Notice that graph $G$, due to its planarity, is 4-colorable. The
graph $G'$, is an extension of graph $G$ that may not be planar,
but preserves 4-Colorability. Now, from Lemma \ref{lem:3col4reg}
and the  $\nit{NP}$-hardness of 3-colorability for planar graphs
with vertices of  degree at most
 4 \cite[theorem 2.3]{gjs-1976}, we obtain

\begin{corollary}\label{cor:color} \em
3-Colorability for regular graphs of  degree 4 (i.e. with all
their vertices of exactly degree 4) is $\nit{NP}$-complete.
\boxtheorem
\end{corollary} We use the construction in Lemma
\ref{lem:3col4reg} as follows: Given any planar graph $G$ of
degree 4, we construct graph $G'$ as in the lemma, which is
regular of degree 4 and 4-colorable. Its 4-colorability is encoded
as a database problem with a fixed set of  first-order
constraints. Since $G'$ is 4-colorable, the database is
consistent. Furthermore, $G'$ uses all the 4 colors available in
the official table of colors, as specified by the ICs. In the
update part, deleting one of the official colors leaves us with
the problem of coloring  graph $G'$ with only the three remaining
colors  (under an A-repair semantics only changes of colors are
allowed to restore consistency), which is possible iff the
original graph $G$ is 3-colorable. Deciding about the latter
problem is $\nit{NP}$-complete \cite{gjs-1976}. We obtain

\begin{theorem}\label{th:incr-set-del-ins} \em For
A-repairs, ground atomic queries, first-order ICs, and update
sequences consisting of tuple deletions, incremental CQA is
\textit{coNP}-hard. \boxtheorem
\end{theorem}
This result applies to first-order ICs and $\nit{delete}$
operations. For incremental CQA in general, update operations that
introduce violations can be in principle of any of the forms
~$\nit{insert, ~delete, ~change}$. Of course, the hardness result
just obtained then trivially applies to general update sequences.

In order to obtain a hardness result for denial constraints (for
which we are assuming update sequences do not contain tuple
deletions), we can use the kind of A-repairs introduced in
\cite{bbfl-2005}.

\begin{theorem}\label{prop:incr-attr-denial} \em Incremental
CQA wrt denial constraints and atomic queries under the wA-repair
semantics is $P^\nit{NP}$-hard. \boxtheorem
\end{theorem}

 Under the attribute-based repairs semantics,
if the update sequence consist of $\nit{change}$ actions, then we
can obtain polynomial time incremental CQA under the additional
condition that the set of attribute values than can be used to
restore consistency is bounded in size, independent from the
database (or its active domain). Such an assumption can be
justified in several applications, like in census-like databases
that are corrected according to inequality-free denial constraints
that force the new values to be taken in the border of a database
independent region
 \cite{bbfl-2005}; and also in applications where denial constraints, this time
containing inequalities, force the attribute values to be taken in
a finite, pre-specified set. The proof is similar to the one of
Proposition \ref{prop:incr-dens}, and the polynomial bound now
also depends on the size of the set of candidate values.

\begin{theorem}\label{prop:bounded} \em
Under  A-repairs that can be obtained using values from a
database-independent bounded set, incremental
  CQA  for first-order boolean queries, denial constraints, and update sequences containing
  only $\nit{change}$ actions is in \textit{PTIME} in the size of the original database.
\boxtheorem
\end{theorem}

\section{Conclusions} \label{sec:concl}
The dynamic scenario for consistent query answering that considers
possible updates on a database
 had not been considered before in
the literature. Doing incremental CQA on the basis of the original
database and the sequence of updates is an important and natural
problem. Developing algorithms that take into account previously
obtained consistent answers that are possible cached and the
updates at hand is a crucial problem for making CQA scale up for
real database applications. Much research is still needed in this
direction.

In this paper we have concentrated mostly on complexity bounds for
this problem under different semantics. When we started obtaining
results for incremental CQA under repairs that differ from the
original instance by a minimum number of tuples, i.e. C-repairs,
we realized that this semantics had not been sufficiently explored
in the literature in the static version of CQA, and that a
comparison was not possible. In the first part of this paper we
studied the complexity of CQA for this semantics. In doing so, we
have developed  graph theoretic techniques that allow us to
connect the certain and possible (or cautious and brave) semantics
for CQA.

Our results show that the incremental complexity is lower than the
static one in several useful cases, but sometimes the complexity
cannot be lowered. It is a subject of ongoing work the development
of concrete and explicit algorithms for incremental CQA. Also the
complexity of incremental CQA under the alternative semantics
presented in Section \ref{sec:other} deserves further
investigation and a more complete picture still has to emerge.

We obtained the first results about fixed parameter tractability
for incremental CQA, where the input, for a fixed database schema,
can be seen as formed by the original database and the update
sequence, whose length is a relevant parameter. This problem
requires additional investigation. It would be interesting to
examine the area of CQA in general from the point of view of
parameterized complexity, and not only the incremental case. For
example, other natural candidates to be a parameter in the
classic, static setting could be: (a) the number of
inconsistencies in the database, (b) the degree of inconsistency,
i.e. the maximum number of violations per database tuple, (c)
complexity of inconsistency, i.e. the length of the longest path
in the conflict graph or hypergraph. These parameters may be
practically significant since in many applications, like census
application \cite{bbfl-2005}, inconsistencies are ``local".

We considered a version of incremental CQA that assumes that the
database is already consistent before updates are executed, a
situation that could have been achieved because no previous
updates violated the given semantic constraints or a repaired
version was chosen before the new updates were executed.

We are currently investigating the dynamic case of CQA in the
frameworks of  \emph{dynamic complexity} \cite{imm-99,Schwentick}
or \emph{incremental complexity} as introduced in
\cite{msvt-1994}. In this case we start with a database $D$ that
is not necessarily consistent -and this is the main new issue
involved- on which a sequence of basic update operations $U_1,
U_2, ..., U_m$ is executed. A clever algorithm for CQA  may create
or update intermediate data structures at each atomic update step,
to help obtain answers at subsequent steps. We are interested in
the computational complexity of CQA after a sequence of updates,
when the data structures created by the query answering algorithm
at previous states are themselves updatable and accessible.

\vspace{2mm} \noindent {\bf Acknowledgments:}~~ Research supported
by NSERC, and EU projects:  Knowledge Web, Interop and Tones. ~L.
Bertossi is Faculty Fellow of  IBM Center for Advanced Studies
(Toronto Lab.). Part of this research was done while L. Bertossi
visited the University of Bolzano during the summer 2005; he
appreciates the hospitality and support of Enrico Franconi and the
KRDB group.

\bibliographystyle{plain}

\section{Appendix: Proofs}\label{sec:proofs}

\defproof{Lemma \ref{lem:graphExt}}{We consider the three cases for membership
of $v$ to maximum
independent sets in $G$. Let $m$ be the cardinality of a maximum
independent set in $G$. We establish now the first bi-conditional.
The second
bi-conditional follows directly from the analysis for the first one. \\
(a) Assume that $v$ belongs to a maximum independent set $I$ of
$G$. In this case, $v'$ can be added to $I$ obtaining an
independent set of $G'$. In this case $|I \cup \{v'\}| \geq m+1$.

Assume that $v$ does not belong a some maximum independent set
$I'$ of $G'$. If $v \notin I'$, then some of its neighbors belong
to $I'$, and then, $v' \notin I'$. In consequence, $I'$ is also a
maximum independent set of $G$. Then, $|I'| = m$. But this is not
possible, because the size of independent set of $I'$ is at least
$m +1$.\\
(b) Assume that $v$ does not belong to any maximum independent
sets of $G$. Then, some of it neighbors can be found in every
maximum independent set of $G$, and none of them can be extended
 with $v'$ to become an independent set of $G'$.

 So, all the maximum independent set of $G$ are maximum independent sets of
 $G'$ of size $m$.

 Assume, that $v$ belongs to all maximum independent sets of $G'$.
  Then none of the neighbors of $v$ can be found in
 independent sets of $G$, and then
$v'$ can be found in all the maximum independent sets of
 $G'$. Since the maximum independent sets of $G'$ have at least
 cardinality $m$, it must hold that the maximum independent sets
 of $G'$ have cardinality at least $m+1$. Then the deleting $v'$
 from all the maximum independent sets of $G'$ will give us
 independent sets of $G$ of size at least $m$, i.e. maximum
 independent sets of $G$. To all of them $v$ belongs. A
 contradiction.}

\vspace{-3mm}
\defproof{Lemma \ref{lem:all2some}}{{\em (sketch)}~ Hang a rhombus from $v$, i.e.
 add three other vertices, two of them connected to $v$, and the
 third one, connected to the two previous ones. Then,
 reason by cases as in the proof of Lemma \ref{lem:graphExt}.}

\vspace{-3mm}
\defproof{Proposition \ref{lem:dec-graphNPlogn}}{For the first claim, given a
graph $G$ and a vertex $v$, build in polynomial time the graph
$G'$ as in Lemma \ref{lem:graphExt}. It holds that $v$ belongs to
some maximum independent set of $G$ iff $v$ belongs to every
maximum independent set of $G'$. Now, $v$ belongs to every maximum
independent set of $G'$ iff $|\mbox{maximum independent set}$
$\mbox{in } G'| - |\mbox{maximum independent set in } G| = 1$.

Since computing the maximum cardinality of a clique can be done in
time $\nit{FP}^\nit{NP(log(n))}$ \cite{krentel} (see also
\cite[theorem 17.6]{papadimitriou94}), computing the maximum
cardinality of an independent set can be done in the same time
(just consider the complement graph). In consequence, in order to
decide about $v$ and $G$, we can compute the cardinalities of the
maximum independent set for $G$ and $G'$ in 2 times
$\nit{FP}^\nit{NP(log(n))}$, and next compute their difference. It
total, we can perform the whole computation in
$\nit{FP}^\nit{NP(log(n))}$. In consequence, by definition of
class $\nit{FP}^\nit{NP(log(n))}$, we can decide by means of a
polynomial time machine that makes $O(\nit{log}(n))$ calls to an
$\nit{NP}$ oracle, i.e. the decision is made in time
$P^\nit{NP(log(n))}$. The same proof works for the second claim.
It can also be obtained from the first claim and Lemma
\ref{lem:all2some}.}

\vspace{-3mm}
\defproof{Corollary \ref{prop:memb}}{Construct the conflict
graph for the instance wrt the FDs. An atomic ground query is
consistently true if the corresponding vertex in the conflict
graph belongs to all the maximum independent sets.  Then use
Proposition \ref{lem:dec-graphNPlogn}.}

\vspace{-3mm}
 \defproof{Proposition \ref{lem:everyGraph}}{Consider a graph $G = \langle V,
E\rangle$, and assume the vertices of $G$ are uniquely labelled.
Consider the database schema with two relations, $\nit{Vertex}(v)$
and $\nit{Edges}(v_1, v_2, e)$, and the denial constraint $\forall
v_1 v_2 \neg(\nit{Vertex}(v_1) \land \nit{Vertex}(v_2) \land
\nit{Edges}(v_1, v_2,e))$. $\nit{Vertex}$ stores the vertices of
$G$. For each edge $\{v_1,v_2\}$ in $G$, $\nit{Edges}$ contains
$n$ tuples of the form $(v_1, v_2,i)$, where $n$ is the number of
vertices in $G$. All the values in the third attribute of
$\nit{Edges}$ are different, say from $1$ to $n |E|$. The size of
the database instance obtained trough this padding of $G$ is still
polynomial in size.

This instance is highly inconsistent, and  its C-repairs are all
obtained by deleting vertices, i.e. elements of $\nit{Vertex}$
alone. In fact, an instance such that all tuples but one in
$\nit{Vertex}$ are deleted, but all tuples in $\nit{Edges}$ are
preserved is a consistent instance. In this case, $n-1$ tuples are
deleted. If we try to achieve a repair by deleting tuples from
$\nit{Edges}$, say $(v_1,v_2,i)$, then in every repair of that
kind all the $n$ tuples of the form $(v_1, v_2,j)$ have to be
deleted as well. This would not be a minimal cardinality repair.

Assume that $I$ is a maximum cardinality independent set of $G$.
The deletion of all tuples $(v)$ from $\nit{Vertex}$, where $v$
does not belong to $I$, is a C-repair. Now, assume that $D$ is a
repair. As we know, only tuples from $\nit{Vertex}$ may be
deleted. Since, in order to satisfy the constraint, no two
vertices in the graph that belong to $D$ are adjacent, the
vertices remaining in $\nit{Vertex}$ form an independent set in
$G$.

In general, the number of deleted tuples is equal to $n-|I|$,
where $I$ is an independent set represented by a repair. So each
minimal cardinality repair corresponds to a maximum independent
set and vice-versa.}

\vspace{-3mm}
\defproof{Corollary \ref{cor:size}}{This follows from Proposition
\ref{lem:everyGraph}, the fact
that C-repairs correspond to maximum cliques in the complement of
the conflict graph \cite{abchrs-2003}, and the
$P^\nit{NP(log(n))}$-completeness of determining the size of a
maximum clique \cite{krentel}.}

\vspace{-3mm}
\defproof{Proposition \ref{prop:membDCs}}{We use the
conflict hypergraph. The problem of determining the maximum clique
size for hypergraphs is in $\nit{FP}^{\nit{NP}(\nit{log}(n)}$ by
the same argument as for conflict graphs: Deciding  if the size of
maximum clique is greater than $k$ is in $\nit{NP}$. So, by asking
a logarithmic number of $\nit{NP}$ queries, we can determine the
size of maximum clique.

The membership to $P^{\nit{NP}(\nit{log}(n))}$ of CQA for the
C-semantics
 still holds for conjunctive queries
without existential variables. In fact, given an inconsistent
database $D$, a query $Q$, and a ground tuple $t$, we check if $t$
is consistent answer to $Q$ from $D$ as follows: Check if $t$ is
an ordinary answer to $Q$ in $D$ (without considering the
constraints). If not, the answer is {\em no}.

Otherwise, let $t_1, \ldots, t_k$ be the database tuples which are
answers to $Q$ in $D$ and produce $t$ as an answer. Since $Q$ does
not contain existential variables, only one such set exists.
Compute the size of a maximum independent set for the graph
representation of $D$, say $m_0$. Compute the size of a maximum
independent set for the graph representation of $D \smallsetminus
\{t_1\}$, say $m_1$. If $m_1 = m_0$, then there exist a maximum
independent set of $D$ that does not contain $t_1$. So, there
exists a minimum repair that does not satisfy that $t$ is an
answer to $Q$. If $m_1 < m_0$, repeat this procedure for all
tuples in $t_1, \ldots, t_k$. Thus, we have to pose $k$ queries
(that is determined only by the size of the query) to an
$\nit{FP}^{\nit{NP(log}(n))}$ oracle.

In consequence, the complexity of CQA for conjunctive queries
without existential variables is in $\nit{P}^{\nit{NP(log}(n))}$.}

\vspace{-3mm}
\defproof{Lemma \ref{lem:block}}{
The new graph $G'$ consists of two copies of $G$, say $G_1, G_2$,
two additional graphs, $I_k, I_{k+1}$, and two extra vertices
$t,b$. Subgraph $I_k$ consists of $k$ mutually disconnected
vertices; subgraph $I_{k+1}$ consists of $k+1$ mutually
disconnected connected vertexes. Each vertex of $G_1$ is adjacent
to each vertex of $I_k$, and each vertex of $G_2$ is adjacent to
each vertex of $I_{k+1}$. Each vertex of $I_k$ is adjacent to $t$,
and each vertex of $I_{k+1}$ is adjacent to $b$.  Finally, $t, b$
are connected by an edge (c.f. Figure 1).

 We claim that vertex $t$ belongs to all maximum independent sets
of $G'$ iff the cardinality of maximum independent set of $G$ is
equal to $k$. To prove this claim, we consider a few, but
representative possible cases. With $I(G)$ we denote an arbitrary
maximum independent set of $G$.
\begin{itemize}
 \item [1.] $|I(G)| < k - 1$: The maximum independent set of $G'$ is
 $I_k \cup I_{k+1}$; with cardinality
 $2k + 1$.
 \item [2.] $|I(G)| = k - 1$: The maximum independent sets of $G'$
 are (a) $I(G_1) \cup I_{k+1} \cup
 \{t\}$, and
 (b) $I_k \cup I_{k+1}$, with cardinality $2k+1$.
 \item [3.] $|I(G)| = k$: The maximum independent set of $G'$ is
 $I_{k+1} \cup I(G_1) \cup \{t\}$, with cardinality $2k+2$.
 \item [4.] $|I(G)| = k + 1$: The maximum independent sets of $G'$ are
 (a) $G_1 \cup G_2 \cup \{t\}$,  (b) $G_1 \cup G_2
 \cup \{b\}$, (c) $G_1 \cup I_{k+1} \cup \{t\}$; with cardinality $2k+3$.
 \item [5.] $|I(G)| > k + 1$: The maximum independent sets of $G'$ are
 (a) $G_1 \cup G_2 \cup \{t\}$, (b) $G_1 \cup G_2 \cup
\{ b\}$;
 with cardinality $2|I| +1$.
\end{itemize}
Only in case $|I(G)| = k$, $t$ belongs to all maximum independent
sets.}

\vspace{-3mm}
\defproof{Proposition \ref{prop:allsets}}{
By reduction from the following
$P^{\nit{NP}(\nit{log}(n))}$-complete decision problem
\cite[theorem 3.5]{krentel}: Given a graph $G$ and an integer $k$,
is the size of a maximum clique in $G$ equivalent to $0~
\nit{mod}~ k$?

Assume graph $G$ has $n$ vertices. We can also assume that $k$ is
not bigger than $n$. Now, we pass to the graph $G'$ that is the
complement of $G$: It has the same vertices as $G$, with every two
distinct vertices being adjacent in $G'$ iff they are not adjacent
in $G$. A maximum independent set of $G'$ is a maximum clique of
$G$ and vice-versa. So, the cardinality of a maximum independent
set of $G'$ is the size of a maximum clique of $G$.

Next, we take advantage of the construction in Lemma
\ref{lem:block} (c.f. Figure 1): For each $m \in \{k, 2k, \cdots,
\lfloor \frac{n}{k} \times k\rfloor \}$, construct the block graph
$B_m(G', t_m)$. (There are $[n/k]$ possible solutions to the
equation $x \equiv 0 ~\nit{mod}~k$.) All these graphs are
disconnected from each other. Next, create a new vertex $t_g$ and
connect it to the vertices $t_m$  of the blocks $B_m(G',t_m)$. It
is easy to check that the resulting graph, say $\overline{G}$, has
its size bounded above by $O(n^4)$.

It holds that vertex $t_g$ does not belong to every maximum
independent set of $\overline{G}$ iff the size of maximum
independent set of $G$ is equivalent to $0\ \nit{mod}\ k$. So, we
have a reduction to the complement of our problem, but the class
$P^{\nit{NP(log}(n))}$ is closed under complement.

In fact, if the size of maximum independent set of $G$ is not
equivalent to $0~ \nit{mod}~k$, then for every block $B$ in
$\overline{G}$, there exists a maximum independent set $I_B$ of
the block $B$ such that $t_B \notin I_B$ ($t_B$ is the top node of
block $B$). The maximum independent set of $\overline{G}$ is
$\{t_g\} \cup \bigcup_B I_B$ (because there are no edges between
blocks and between $t_g$ and other vertices besides $t_B$).
Consider any independent set $I$ of $\overline{G}$ that does not
contain $t_g$. The size of the projection of $I$ on any block is
not greater than the size of the maximum independent set of the
block; so $|I| \leq |\bigcup_B I_B|$. So, $t_g$ belongs to every
maximum independent set of $\overline{G}$.

Now, if the size of a maximum independent  set  of $G$ is
equivalent to $0\ \nit{mod}\ k$, then there exists one block
$B_{\!o}$ such that $t_{B_o}$ belongs to every maximum independent
set $I_{B_o}$ of $B_{\!o}$, while for all other blocks $B$ there
exists $I_B$ such that $t_B \notin I_B$. Consider a maximum
independent set $I_t$ of $\overline{G}$ that contains $t_G$.

Every maximum independent set of $\overline{G}$ that contains
$t_g$ is of the form $\{t_g\}$ union of maximum independent sets
from the blocks $B$ other than $B_o$ that do not contain their
corresponding $t_B$ union any maximum independent set of $B_{\!o}
\smallsetminus \{t_{B_o}\}$. The size of such a set is $s = 1 +
\sum_{B \neq B_o} |I(B)| + (|I_{B_o}| -1)$. A maximum independent
set $I$ that does not contain $t_g$, is the union of maximum
independent sets $I_B$ of all the blocks $B$ of $\overline{G}$,
and its size is equal to $\sum_B |I_B|$, i.e. $s$. Then, there
exists a maximum independent set that does not contain $t_g$. }

\vspace{-3mm}
\defproof{Theorem \ref{teo:hardness}}{Membership follows from
Proposition \ref{prop:membDCs}. Now we prove hardness.
 For a graph $G$ and integer $k$, we
construct a database $D$, such that the consistent answer to a
ground atomic query $Q$ can be used to decide if the size of a
maximum clique of $G$ is equivalent to $0~ \nit{mod}~ k$ (c.f.
proof of Proposition \ref{prop:allsets}). Construct the graph
$\overline{G}$ as in Proposition \ref{prop:allsets}. Encode graph
$\overline{G}$ as a database inconsistency problem, introducing a
unary relation $V$ (for vertices) and $E$ (3-ary), where $E$
corresponds to the edge relation in $\overline{G}$ plus a third
padding attribute to make changing it more costly. For each vertex
$v \in \overline{G}$, there is a tuple $(v)$ in $V$.

We also introduce the denial constraint: $\forall v_1 \forall v_2
\neg(V(v_1) \wedge V(v_2) \wedge E(v_1, v_2,\!\_))$ (an underscore
means any variable implicitly universally quantified).
 For each edge
$\{v_1, v_2\} \in \overline{G}$, create $n$ different versions
$(v_1, v_2,p)$ in $E$, as in the proof of Proposition
\ref{lem:everyGraph}. The effect of fixing the database wrt the
given denial constraint may be the removal of tuples representing
vertices or/and the removal of  tuples representing edges. We want
to forbid the latter alternative because those repairs do not
represent maximum independent set; and this is achieved by making
them more expensive than vertex removal through the padding
process.

The consistent answer to the query $V(t_g)$ is {\em no}, i.e. not
true in all repairs, iff $t_g$ does not belong to all maximum
independent sets of $\overline{G}$ iff the size of a maximum
independent set of $G'$ is equivalent to $0~ \nit{mod}~ k$ iff the
size of a maximum clique of $G$ is equivalent to $0~ \nit{mod}~
k$.}

\vspace{-3mm}
\defproof{Proposition \ref{theo:weightedpro}}{Membership is proved as for the
C-repair semantics (c.f.  Proposition \ref{prop:membDCs}): Repairs
correspond to maximum weighted independent sets of the associated
hypergraph $G$. The weight of a maximum weighted independent set
can be found in $P^\nit{NP}$ (as for  independent set, but
$\nit{log}(O(2^n)) = \nit{poly}(n)$ oracle calls are required). To
check if a vertex $v$ belongs to all maximum weighted independent
sets, it is good enough to compute weights of maximum independent
sets for $G$ and $G \setminus \{v\}$.}

\vspace{-3mm}
\defproof{Proposition \ref{prop:incr-dens}}{
For denial constraints tuple deletions do not introduce any
violations, so we consider a sequence $U$ consisting of tuple
insertion and updates.

 Assume that $k$ of the $m$ inserted
tuples violate ICs, perhaps together with some tuples already in
$D$. If we delete $k$ violating tuples, then we get a consistent
database $D'$;  so a minimal repair is at a distance less than or
equal to $k$ from $D$. To find all minimal repairs it is good
enough to check no more than $N = {\small \left(
\begin{array}{c} n + m \\ 1 \end{array} \right) + \left(
\begin{array}{c} n + m\\ 2
\end{array} \right) + \cdots\ + \left( \begin{array}{c} n + m\\ k
\end{array} \right)}$
 repairs, where $|D| = n$. If $m$ is small, say less than $c
\cdot n$, then $N < k {\small \left( \begin{array}{c} n + m\\
k\end{array}\right)}  \leq m {\small \left( \begin{array}{c} n\\
m\end{array}\right)}^{\!\!m} <  mn^m$. Thus, the incremental
complexity of the CQA is polynomial wrt $n$.

In case $U$ contains change updates, the proof is essentially the
same, but the role of $m$ is taken by $m \cdot a$, where $a$ is
the maximum arity of the relations involved. This is because we
have to consider possible changes in different attributes.}

\vspace{-3mm}
\defproof{Proposition \ref{prop:fpt}}{
First, it is known that the problem of, given a graph $G$ and a
number $k$, determining if there exists a vertex cover of size
less than or equal to $k$ is in FPT \cite{downfel}. We will use
this problem to solve ours.

Now, let us assume that we have a consistent database $D$ of size
$n$, and we update it inserting $k$ new tuples, obtaining an
inconsistent database $D'$ with conflict graph $G$. The size of
$G$ is $O(n)$ by our assumption on the size of $m$ in comparison
with $n$. Every C-repair of $D'$ is a maximum independent set of
$G$, and can be obtained by deleting from $G$ a minimum vertex
cover, because the problems are complementary. So, a minimum
vertex cover corresponds to the vertices that are to be deleted to
obtain a repair.

Since the original database $D$ is consistent, the vertices of $G$
corresponding to database tuples in $D$ are all disconnected from
each other. In consequence, edges may appear only by the update
sequence, namely between the $m$ new tuples or between them and
the elements of $D$. Then, we know that there is a vertex cover
for $G$ of size $m$. However, we  do not know if it is minimum.

In order to find the size of a minimum vertex cover of $G$, we may
start doing binary search from $m$, applying an FPT algorithm for
vertex cover.  Each check for vertex cover, say for value $m_i$,
can be done in $O(1.2852^m_i + m_i \cdot n)$ \cite{chen}. Then
$\nit{log}(m)$ checks take time $O(\nit{log}(m) \cdot (1.2852^m +
m \cdot n)) \leq O(f(m) \cdot n)$, with $f$ an exponential
function in $m$. So, it is in $\nit{FPT}$ obtaining the size of a
minimum vertex cover for $G$, which gives us the minimum number of
tuples to remove to restore consistency.

Now, for CQA we want to check if a vertex $R(\bar{t})$ belongs to
all maximum independent sets of $G$, which happens if it does not
belong to any minimum vertex covers. This can be determined by
checking the size of minimum vertex cover for $G'$ and $G'
\smallsetminus \{R(\bar{t})\}$. If they are the same, then
$R(\bar{t})$ belongs to all maximum independent sets and the
consistent answer to the query $R(\bar{t})$ is $\nit{yes}$.}

\vspace{-3mm}
\defproof{Proposition \ref{lem:param-compl-incr}}{
 By uniform
reduction from the $\nit{MONOTONE~W[1]}$-hard problem
\cite{df-1995} \emph{WEIGHTED MONOTONE 3CNF SAT}, which is defined
as follows: Given a 3CNF monotone circuit $C$ and an integer $k$,
is it possible to make exactly $k$ of the inputs $1$ and obtain
output $1$ for $C$?

The database schema consists of relations $\nit{Clause}(C, V_1,
V_2, V_3, p)$, $\nit{Var}(V)$, $\nit{Cond}(X,Y)$, where $p$ and
$Y$ are dummy variables intended to create many copies of a tuple,
to forbid the deletion of those tuples by making the potential
repair too costly.  The integrity constraint is~ $\forall C V_1
V_2 V_3 p y \neg (Clause(C, V_1, V_2, V_3, p) \land Var(V_1) \land
Var(V_2) \land Var(V_3) \land \nit{Cond}(1,y))$. Given a monotone
3CNF formula $\Psi = \psi_1 \land \psi_2 \land \cdots \land
\psi_m$ and a parameter $k$, for each clause $\psi_i = (x_{i_1}
\lor x_{i_2} \lor x_{i_3})$, where the $x_{i_j}$ are atoms,
 store in $\nit{Clause}$ $n$ copies of the form $(i, x_{i_1}, x_{i_2},
 x_{i_3},p)$
(replace variable by any new constant if a clause has less than
three variables). For each variable $x$ in $\Psi$, store $x$ in
$Var$. $C$ is initially empty. The resulting database is
consistent.

Now, on the update part, insert $(1, i)$ into $\nit{Cond}$, $i =
1, \ldots k$. Then there exists an assignment with weight less
than $k$ iff $\nit{Cond}(1,1)$ is $\nit{false}$ in every repair.

Since we have to determine if there exists a satisfying assignment
with weight  exactly $k$, it is good enough to ask a query to two
databases, built as before, but for both $k$ and $k+1$, which is
compatible with the definition of \emph{parametric reduction},
that allows to use of a constant number of instances. In our case,
since we have that: (a) if weight $< k$, then consistent answer is
{\em yes}, (b) if weight is equal to $k$, then the consistent
answer is \emph{false} (i.e. false in all repairs), and (c) if
weight $> k$, the consistent answer is \emph{false}. So, we
construct two instances, for $k$ and $k+1$. The weight is equal to
$k$ iff the consistent answer for the first instance is
\emph{false} and for the second one it is {\em yes}. }

\vspace{-3mm}
\defproof{Proposition \ref{lem:incr-icd-tds-msd}}{By reduction from static
  CQA for (existentially quantified) conjunctive queries and denial ICs under
  minimal set semantics, which is \textit{coNP}-hard
  \cite{cm-2005}. Consider an instance for this problem consisting of a database
  $D$, a set of denial ICs $\IC$, and a query $Q$.

  For every denial $\nit{ic} \in \IC$, pick up a relation $R^\nit{ic}$ in it and
  expand it to a relation
  $\overline{R^\nit{ic}}$ with an extra attribute $\nit{Control}$.
  Also add a new, one attribute relation
  $\nit{Controler}(A)$. Next, transform each integrity constraint
$\nit{ic}\!: ~
  \forall \bar{x} \neg (P(\bar{x}) \land \cdots \wedge R^\nit{ic}(\bar{x}) \land
  \cdots \wedge \gamma)$ into
  $\nit{ic}^{\prime}\!: ~\forall \bar{x}\forall \nit{contr} \neg (P(\bar{x}) \land
  \cdots \wedge
  \overline{R^\nit{ic}}(\bar{x},
  \nit{contr}) \land \nit{Controler(contr)} \land \gamma)$. We
  obtain a set $\IC'$ of denial constraints. The original database
  $D$ is extended to a database $\overline{D}$ with the new relation
  $\nit{Controler}$, which is initially empty, and the relations $\overline{R^\nit{ic}}$,
  whose extra attributes $\nit{Contr}$ initially take all
the value $1$. Due to the extension of $\nit{Controler}$, $\IC'$
is satisfied.

  Now in the incremental context, we consider the inconsistent instance $\overline{D}'$
  obtained via the update $\nit{insert}(\nit{Controler}(1))$ on $\overline{D}$.
The S-repairs  of $\overline{D}'$ wrt $\IC'$ are: (a)
$\overline{D}$ and (b) all the S-repairs
  of $\overline{D}$ (plus the tuple $\nit{Controler}(1)$ in each of them), which are in one-to-one
  correspondence with the S-repairs of $D$ wrt $\IC$.
Now, for a conjunctive query $Q$ in the language of $D$, produce
the conjunctive query
  $Q'\!: \exists \cdots y_\nit{ic} \cdots Q\frac{\cdots R^\nit{ic}(\bar{x}) \cdots}{\vspace*{1mm}\cdots
  \overline{R^\nit{ic}}(\bar{x},y_\nit{ic}) \cdots}$ in the
  language of $\overline{D}$,\footnote{$E\frac{E_1}{E_2}$ means the expression obtained by replacing
  in expression $E$ the subexpression $E_1$ by expression $E_2$.} where each atom
  $R^\nit{ic}(\bar{x}))$ in $Q$ is replaced by $\exists y_\nit{ic}
  \overline{R^\nit{ic}}(\bar{x},y_\nit{ic})$.

  Notice that all the repairs in (b) are essentially contained
 in $\overline{D}$, except for the tuple $\nit{Controler}(1)$, whose
  predicate does not appear in the queries. This is because denial
  constraints are obtained by tuple deletions. In consequence, any
  answer to the conjunctive (and then monotone) query in a repair
  in (b) is also an answer in the repair in (a). In consequence,
  the repair $\overline{D}$ does not contribute with any new
  consistent answers, neither invalidates any answers obtained by
  the repairs in (b). So, it holds $\nit{Cqa}(Q,D,\IC) =
  \nit{Cqa}(Q',\overline{D}',\IC')$.}

\vspace{-3mm}
  \defproof{Theorem \ref{theo:wmccs}}{We can adapt the proof of theorem 4 in
\cite{bbfl-2005} about the $\Delta_2^P$-hardness of CQA under
minimum square distance. We provide a $\nit{LOGSPACE}$-reduction
from the following problem \cite[theorem 3.4]{krentel}: Given a
Boolean formula $\psi(X_1, \cdots, X_n)$ in 3CNF, decide if the
last variable $X_n$ is equal to $1$ in the lexicographically
maximum satisfying assignment (the answer is ${\it No}$ if $\psi$
is not satisfiable).

Create a database schema with relations:~ $\nit{Clause(id,
Var}_1,$ $ \nit{Val}_1, \nit{Var}_2, \nit{Val}_2,$ $\nit{Var}_3,
\nit{Val}_3)$, $\nit{Var(var, val)}$, $\nit{Dummy}(x)$, with
denial constraints:\\
$\forall var, val  \neg (Var(var, val) \land val \not = 0  \land
val \not = 1)$,\\
$\forall id, v_1, x_1, v_2, x_2, v_3, x_3 \neg (Cl(id, v_1, x_1,
v_2, x_2, v_3, x_3)
 \land Var(\_, v_1, x'_1)  \land Var(\_, v_2, x'_2)$
 $\land~ Var(\_, v_3, x'_3) ~\land~
 x_1 \not = x'_1  \land x_2 \not = x'_2 \land x_3 \not =  x'_3
\land
 \nit{Dummy}(1))$.

\noindent The last denial can be replaced by 8 denial constraints
without inequalities considering all the combination of values for
$x_1, x_2, x_3$ in $\{0,1\}$.

Assume now that $C_1, \ldots, C_m$ are the clauses in $\psi$. For
each propositional variable $X_i$ store in table $\nit{Var}$ the
tuple $(X_i, 0)$, with weight $1$, and $(X_i,1)$ with weight
$2^{n-i}$. Store tuple $1$ in $Dummy$ with weight $2^n\times 2$.
For each clause $C_i = l_{i_1} \lor l_{i_2} \lor l_{i_3}$, store
in $\nit{Clause}$  the tuple $(C_i, X_{i_1}, \tilde{l}_{i_1},
X_{i_2}, \tilde{l}_{i_2}, X_{i_3}, \tilde{l}_{i_3})$, where
$\tilde{l}_{i_j}$ is equal to $1$ in case of positive occurrence
of variable $X_{i_j}$ in $C_i$; and  to $0$, otherwise. For
example, for $C_6 = X_6 \lor \neg X_9 \lor X_{12}$, we store
$(C_6, X_6,1, X_9, 0,  X_{12}, 1)$. The weight of this tuple is
$2^n$.

Then the answer to the ground atomic query $\nit{Var}(X_i,1)$ is
$\nit{yes}$ iff the variable $X_i$ is assigned value $1$ in the
lexicographically maximum assignment (in case  such a satisfying
assignment exists). In case a satisfying assignment does not
exist, then the tuple in $\nit{Dummy}$ has to be changed in order
to satisfy the constraints. No attribute value in a tuple in
$\nit{Clause}$ is changed, because the cost of such a change is
higher than a change in the $\nit{Dummy}$ relation.}

\vspace{-3mm}
\defproof{Lemma \ref{lem:3col4reg}}{If a vertex $v$ in $G$ has degree 2, then
we transform it into a vertex of degree 4 by hanging from it an
``ear" as shown in the figure, which is composed of three
connected versions of the graph $H_3$ \cite[Theorem 2.3]{gjs-1976}
plus two interconnected versions of a box graph (c.f. figure
below).

It is easy to see that the ``ear" is regular of degree 4, is
3-colorable (as shown in the picture
  with colors {\tt r,g,b}), but not
  planar. Hanging the ear adds a constant number of vertices. Now
  we have to deal with the set $V_\nit{odd}$ of vertices of degree 1 or 3 (vertices of degree 0 can be
  ignored). By Euler's theorem, $V_\nit{odd}$ has an even
  cardinality. This makes it possible to pick up
disjoint pairs $\{v_1, v_2\}$ of elements

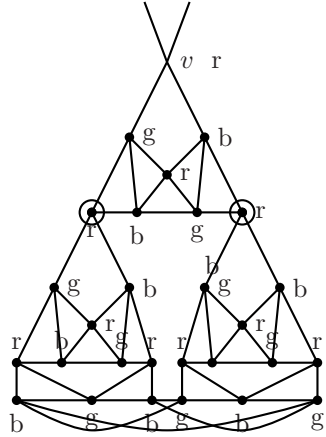
\begin{multicols}{2}
\psset{xunit=1mm,yunit=1mm}
\begin{pspicture}(-5,-8)(55,60) \psline(25,50)(28,58)
\psline(25,50)(22,58) %\pscurve(25,53)(0,2)(25,0)(50,2) \psdot(25,50)
\uput[r](25,50){$v$~~r} \psdot(20,40) \uput[r](20,40){g}
\psdot(30,40) \uput[r](30,40){b} \psdot(25,35) \uput[r](25,35){r}
\psdot(15,30) \uput[d](15,30){r} \psdot(21,30) \uput[d](21,30){b}
\psdot(29,30) \uput[d](29,30){g} \psdot(35,30) \uput[r](35,30){r}
\pscircle(15,30){5pt} \pscircle(35,30){5pt} \psline(25,50)(20,40)
\psline(25,50)(30,40) \psline(20,40)(25,35) \psline(25,35)(30,40)
\psline(20,40)(15,30) \psline(20,40)(21,30) \psline(25,35)(21,30)
\psline(25,35)(29,30) \psline(30,40)(29,30) \psline(30,40)(35,30)
\psline(15,30)(21,30) \psline(21,30)(29,30) \psline(29,30)(35,30)
\psdot(15,30) \psdot(10,20) \uput[r](10,20){g} \psdot(20,20)
\uput[r](20,20){b} \psdot(15,15) \uput[r](15,15){r} \psdot(5,10)
\uput[u](5,10){r} \psdot(11,10) \uput[u](11,10){b} \psdot(19,10)
\uput[u](19,10){g} \psdot(23,10) \uput[u](23,10){r}
\psline(15,30)(10,20) \psline(15,30)(20,20) \psline(10,20)(15,15)
\psline(15,15)(20,20) \psline(10,20)(5,10) \psline(10,20)(11,10)
\psline(15,15)(11,10) \psline(15,15)(19,10) \psline(20,20)(19,10)
\psline(20,20)(23,10) \psline(5,10)(11,10) \psline(11,10)(19,10)
\psline(19,10)(23,10) \psdot(35,30) \psdot(30,20)
\uput[r](30,20){g} \psdot(40,20) \uput[r](40,20){b} \psdot(35,15)
\uput[r](35,15){r} \psdot(27,10) \uput[u](27,10){r} \psdot(31,10)
\uput[u](31,20){b} \psdot(39,10) \uput[u](39,10){g} \psdot(45,10)
\uput[u](45,10){r} \psline(35,30)(30,20) \psline(35,30)(40,20)
\psline(30,20)(35,15) \psline(35,15)(40,20) \psline(30,20)(27,10)
\psline(30,20)(31,10) \psline(35,15)(31,10) \psline(35,15)(39,10)
\psline(40,20)(39,10) \psline(40,20)(45,10) \psline(27,10)(31,10)
\psline(31,10)(39,10) \psline(39,10)(45,10) \psdot(5,5)
\uput[d](5,5){b} \psdot(15,5) \uput[d](15,5){g} \psdot(23,5)
\uput[d](23,5){b} \psline(5,5)(15,5) \psline(15,5)(23,5)
\psline(15,5)(5,10) \psline(15,5)(23,10) \psline(5,5)(5,10)
\psline(23,5)(23,10) \psdot(27,5) \uput[d](27,5){g} \psdot(35,5)
\uput[d](35,5){b} \psdot(45,5) \uput[d](45,5){g}
\psline(27,5)(35,5) \psline(35,5)(45,5) \psline(35,5)(27,10)
\psline(35,5)(45,10) \psline(27,5)(27,10) \psline(45,5)(45,10)
\pscurve(5,5)(15,1)(27,5) \pscurve(5,5)(25,1)(45,5)
\psline(23,5)(27,5) \pscurve(23,5)(35,1)(45,5)
\end{pspicture}

\noindent
  of $V_\nit{odd}$, leaving every
vertex coupled to some other vertex.
 For each such pair, $\{v_1,
v_2\}$, add an extra vertex $v'$ connected to (only) $v_1$ and
$v_2$. This trio is 3-colorable. Now $v_1, v_2$ have degree 2 or
4. ~~From those that become of degree 2, hang the ``ear" as
before. In this way, all the nodes become of degree 4. The number
of added vertices is polynomial in the size of the original graph.
The 4-colorability of $G'$ follows from the 4-colorability of $G$
(every planar graph is 4-colorable) and the 4-colorability of the
hanging ears.}
\end{multicols}

 \defproof{Corollary \ref{cor:color}}{From Lemma \ref{lem:3col4reg} and the
  $\nit{NP}$-hardness of 3-colorability for planar graphs with vertices of
  degree at most
  4 \cite{gjs-1976}.}

\vspace{-3mm}
  \defproof{Theorem \ref{th:incr-set-del-ins}}{If the update operation $U$ is a
  $\nit{delete}$ of
  a database atom, we reduce to our problem
  3-Colorability of
  planar graphs $G$ with vertex degree at most 4, which is
  \textit{NP}-complete \cite{gjs-1976}. Given such a non-empty graph $G$, we construct
  graph $G'$ as in
  Lemma \ref{lem:3col4reg}, which is also 4-colorable (because $G$ is and the
ears too).

Let  $E(X, Y)$ be a database relation encoding the edges of the
graph, $\nit{Coloring}$ a 2-ary database relation storing a
coloring of the vertices, and $\nit{Colors}$ a unary relation
storing the four colors
  allowed. Notice that a 4-coloring of $G$ can be found in polynomial time
  \cite{rsst-96}. Then also a 4-coloring for $G'$ can be found in polynomial
time (a 4-coloring for the ears can be given once and for all). The ICs,
essentially denials and inclusion dependencies, are as
  follows:
  \begin{enumerate}
  \item Every node is colored:~ $\forall xy \exists z (E(x,y) \rightarrow
  \nit{Coloring}(x, z))$.
  \item
  Nodes have one color: $\forall x y_1 y_2 \neg(\nit{Coloring}(x, y_1) \land
  \nit{Coloring}(x, y_2) \land y_1 \not = y_2)$.

  \item
  Colors must be allowed:~
~$\forall xy (\nit{Coloring}(x,y) \rightarrow \nit{Colors}(y))$.

  \item
  Vertex degree is not less than 4:~
  $\forall x ( \exists y E(x,y) \rightarrow \exists y_1 y_2 y_3 y_4 (E(x,y_1)
  \land E(x,
  y_2) \land E(x, y_3) \land E(x, y_4) \land y_1 \not = y_2
  \land y_1 \not = y_3 \land y_1 \not = y_4 \land y_2 \not = y_3 \land
  y_2 \not = y_3 \land y_3 \not = y_4))$.

  \item
  Vertex degree is not bigger than 5:~
  $\forall x y_1 \cdots y_5 \neg (E(x, y_1) \land \cdots \land
  E(x,y_5) \land y_1 \not = y_2 \cdots \land y_4 \not =  y_5)$.

  \item
  Only vertices are colored:~
  $\forall xy\exists z(\nit{Coloring}(x,y) \rightarrow E(x,z))$.

  \item All colors are used:~
  $\forall x\exists z (\nit{Colors}(x) \rightarrow \nit{Coloring}(z,x))$.
  \item $E$ is symmetric:~ $\forall xy(E(x,y) \rightarrow E(y,x))$.
  \item Adjacent vertices have different colors:\\ $\forall
  xyuw \neg (E(x,y) \wedge \nit{Coloring}(x,u) \wedge
  \nit{Coloring}(y,w)
\wedge u = w)$.
  \end{enumerate}
  The initial database $D$ stores the graph  $G'$, together with its
  4-coloring (that does use all 4 colors). This is a consistent instance.

  For the incremental part, if the update $U$ is the deletion of a color,
  e.g. $\nit{delete}_\nit{Colors}(c)$, i.e. of
  tuple $(c)$ from $\nit{Colors}$,  the instance becomes inconsistent, because an
inadmissible color  is being used in the coloring. Since repairs
can be obtained by changing attribute values in existing tuples
only, the only possible
  repairs are the 3-colorings of
  $G'$ with the 3 remaining colors  (if such colorings exist), which are obtained
  by changing colors in the second attribute of $\nit{Coloring}$.
  If there are no colorings, there are no repairs.

The query $Q\!:~ \nit{Colors}(c)?$ is consistently true only in
case there is no 3-coloring of the original graph $G$, because it
is true in the empty set of repairs.}

\vspace{-3mm}
\defproof{Theorem \ref{prop:incr-attr-denial}}{
We reason basically as in the proof of theorem 4(c) in
\cite{bbfl-2005}; just introduce a new relation $\nit{Dummy}$, and
transform every denial $\forall \bar{y} \neg (A_1 \wedge \cdots
\wedge A_s)$ there into $\forall \bar{y} \forall x \neg (A_1
\wedge \cdots \wedge A_s \wedge \nit{Dummy}(x))$.
 If we start with the empty extension for $\nit{Dummy}$, the
database is consistent. On the update part, if we insert the tuple
$\nit{Dummy}(c)$ into the database, and the original denials were
inconsistent in the given instance, then we cannot delete that
tuple and no change in it can repair any violations. Thus, the
only way to repair database is  as in \cite{bbfl-2005}, which
makes CQA $P^\nit{NP}$-hard.}

\end{document}